# Ultrafast nematic-orbital excitation in FeSe

T. Shimojima[1,2,*], Y. Suzuki[2], A. Nakamura[1,2], N. Mitsuishi[2], S. Kasahara[3], T. Shibauchi[4], Y. Matsuda[3], Y. Ishida[5], S. Shin[5] and K. Ishizaka[1,2]

[1]RIKEN Center for Emergent Matter Science (CEMS), Wako 351-0198, Japan

[2]Quantum-Phase Electronics Center (QPEC) and Department of Applied Physics, The University of Tokyo, Tokyo 113-8656, Japan

[3]Department of Physics, Kyoto University, Kyoto 606-8502, Japan

[4]Department of Advanced Materials Science, The University of Tokyo, Kashiwa, 277-8561, Japan

[5]Institute for Solid State Physics (ISSP), The University of Tokyo, Kashiwa, 277-8581, Japan.

*Correspondence to: takahiro.shimojima@riken.jp

**The electronic nematic phase is an unconventional state of matter that spontaneously breaks the rotational symmetry of electrons. In iron-pnictides/chalcogenides and cuprates, the nematic ordering and fluctuations have been suggested to have as-yet-unconfirmed roles in superconductivity. However, most studies have been conducted in thermal equilibrium, where the dynamical property and excitation can be masked by the coupling with the lattice. Here we use femtosecond optical pulse to perturb the electronic nematic order in FeSe. Through time-, energy-, momentum- and orbital-resolved photo-emission spectroscopy, we detect the ultrafast dynamics of electronic nematicity. In the strong-excitation regime, through the observation of Fermi surface anisotropy, we find a quick disappearance of the nematicity followed by a heavily-damped**

**oscillation. This short-life nematicity oscillation is seemingly related to the imbalance of Fe $3d_{xz}$ and $d_{yz}$ orbitals. These phenomena show critical behavior as a function of pump fluence. Our real-time observations reveal the nature of the electronic nematic excitation instantly decoupled from the underlying lattice.**

Iron-based superconductors exhibit attractive properties represented by high-transition-temperature ($T_c$) superconductivity and comlex competing phases[1]. Their electronic strucrtures are consisting of multiple iron $3d$ orbitals, thus giving rise to a variety of antiferroic and ferroic ordering phenomena involving spin and orbital profiles[2-4]. Among these mysterious phases, there has been increasing interest in the nematic order[5-13] which spontaneously breaks the rotatioal symmetry of electrons and triggers the lattice instability[8]. Recent investigations by electronic Raman scattering[14,15] and elastoresistivity measurements[11] unveiled the fluctuation of the electronic nematicity and its critical behavior commonly in the optimally doped regimes of different material families[11].

FeSe exhibits superconductivity ($T_c$ = 9 K) and a nematic order accompanied by the tetragonal-to-orthorhombic lattice deformation ($T_s$ = 90 K) without any magnetic order[16]. Its electronic structure is given in Fig. 1 (Supplementary Note 1). In the tetragonal phase, FeSe exhibits a circular Fermi surface around the Γ point (Fig. 1a). Along $k_x$ ($k_y$), the hole band forming the Fermi surface has the $yz$ ($xz$) orbital component. Note that such momentum ($k$)-dependent orbital characters keep the four-fold ($C_4$) symmetry (Fig. 1b). In the orthorhombic (nematic) phase, the $k$-dependent orbital polarization[17] modifies the Fermi surface shape into an elliptical one (nematic Fermi surface) as shown in Fig. 1c, resulting in inequivalent Fermi momenta ($k_F$) along $k_x$ and $k_y$ ($k_{Fx} < k_{Fy}$). At the same time, the orbital components at the $k_F$'s are mixed, especially along $k_x$ as shown in Fig. 1d (Supplementary Note 2). While these characteristics associated with the nematic order have been verified through intensive angle-resolved photo-emission spectroscopy (ARPES) studies[17-20], the understanding of dynamics and excitations peculiar to this condensed state are yet lacking.

Time-resolved ARPES (TARPES) has potential impact of resolving electron dynamics not only in energy and momentum but also into spin and orbital degrees of freedoms. A wide range of materials has been investigated for clarifying the electronic dynamics, such as recombination of the superconducting quasiparticles[21,22], fluctuating charge density wave[23], collapse of long-range order[24,25] and coupling with optical phonons[25-27]. These results, which are inaccessible from equilibrium state, contributed to the deeper understanding of the novel quantum states, especially of short lifetime. Here we use TARPES to track the ultrafast dynamics of the electronic nematicity in FeSe. By combining the detwinned crystals with linear-polarized probe laser, we can selectively obtain the electrons of *xz* and *yz* orbitals (Supplementary Note 3). With this TARPES setup[28] (Fig. 1e), the ultrafast dynamics of the nematic Fermi surface and the orbital-dependent carrier dynamics can be visualized.

**Results**

**Ultrafast dynamics of nematic Fermi surface.**

Immediately after the photo-excitation ($t = 120$ fs), the hole bands around the $\Gamma$ point along $k_x$ and $k_y$ exhibit remarkable momentum shifts with the opposite signs, and take the comparable $k_F$ values as indicated by the red and blue arrows in Fig. 2a,b. This observation suggests that the elliptical Fermi surface quickly changes to circular by the photo-excitation, thus indicating the melting of the nematic order. Figure 2c displays the fluence ($F$)-dependence, where the shift of $k_{Fy}$ at $t = 120$ fs ($\Delta k_{Fy}$) gradually increases as a function of $F$ (weak-excitation regime), and saturates at $F > F_c = \sim 200$ µJcm$^{-2}$ where the isotropic Fermi surface is attained (strong-excitation regime).

Here we track the time dependence of $\Delta k_{Fy}(t)$ for respective $F$ (Fig. 2d). As shown in Fig. 2c,

$\Delta k_{Fy}(t)$ indicates the sudden decrease at $t \sim 120$ fs representing the melting of nematicity, followed by the subsequent recovery in ~1 ps. The overall picture of this transient Fermi surface for $F > F_c$ is shown in Fig. 2e. We further find that the recovery clearly becomes faster for $F > F_c$, and some modulated feature appears. The time dependences of $k_{Fx}$ and $k_{Fy}$ for 220 μJcm$^{-2}$, where the modulation appears most strongly, are presented in Fig. 2f. These data indicate that the $C_2$ anisotropy in $k_F$ is completely suppressed ($k_{Fx} \approx k_{Fy}$) within the time resolution (250 fs), followed by an anomalous hump in the recovery. These behaviors of $k_{Fx}$ and $k_{Fy}$ can be reproduced by the functions including the damped oscillation term in the form of $k_F(t) = k_{F0} + k_{F1}\exp(-t/\tau_1) + k_{F2}\exp(-t/\tau_2) + k_{F3}\exp(-t/\tau_3)\cos(2\pi t/t_p)$ convoluted by the Gaussian of the time resolution, with common values of $\tau_1 = 830 \pm 50$ fs, $\tau_2 > 80$ ps, $\tau_3 = 550 \pm 50$ fs, and $t_p = 1.4 \pm 0.05$ ps. The observed anti-phase oscillation of $k_{Fx}$ and $k_{Fy}$ directly represents the Pomeranchuk-type oscillation of Fermi surface[29], being intensively discussed as the fundamental excitation in the electronic nematic state. The time scale of the oscillatory response (1.4 ps, 3.1 meV) is considerably slow as compared to the coherent $A_{1g}$ optical phonon (190 fs, 22 meV) which is known to strongly couple to the electronic state in this system[25-27]. Their possible interplay is unfortunately hidden in the present TARPES data, possibly due to the duration of the pump pulse (170 fs) comparable to the time period of $A_{1g}$ mode (190 fs) that tends to vanish the coherent oscillation.

**Orbital-dependent carrier dynamics.**

Based on the behavior of transient Fermi surface, we focus on the orbital-dependent carrier dynamics. With $p$-polarized probe pulse, we obtain the energy-distribution curves (EDCs) for $xz$ and $yz$ electrons around the Γ point by integrating $k_y$ and $k_x$ in $0.00 \pm 0.04$ Å$^{-1}$, respectively (see Supplementary Note 3 for experimental settings). The main peak of EDC around −18 meV in Fig. 3a,b,e,f corresponds to the top of the middle (β) hole band sinking below $E_F$, predominantly of $yz$ orbital character (Fig. 1d). In the weak-excitation regime of $F = 40$ μJcm$^{-2}$ (Fig. 3a,b), the main peak gets rapidly suppressed, and electrons are excited toward the unoccupied state. We note that the excited tail intensity of EDCs at $E > E_F$ is very low for $yz$, being consistent with the predominantly $xz$ character of the outer ($\alpha$) hole band top at $\Gamma$ (Fig. 1d)[30]. Here we track the evolution of the corresponding photo-electron intensities $\Delta I(t)$ at $E - E_F = 7.5 \pm 2.5$ meV and $k = 0.00 \pm 0.04$ Å$^{-1}$, i.e. black rectangles in Fig. 1b,d. Figures 3c,d show that $\Delta I(t)$ for $xz$ and $yz$ exhibit the similar exponential decay function with two time constants 850 fs and >80 ps, thus indicating the mostly equivalent relaxation processes of both orbitals.

In the strong-excitation regime, the photo-response changes drastically. The EDCs in Fig. 3e,f show that the photo-excited states at $E > E_F$ also appear in $yz$, indicating that the $C_4$ isotropic state (Fig. 1b) is achieved by the strong photo-excitation ($F = 220$ μJcm$^{-2}$). On the other hand, the excited intensity of $xz$ shows a nonmonotonic relaxation which keeps increasing from $t = 120$ fs to 700 fs as indicated by the black arrow in Fig. 3e, being markedly different from $yz$. As shown in Fig. 3g,h, $\Delta I(t)$ of $xz$ exhibits the retarded maximum at $t_{\text{ret}} = \sim 700$ fs, whereas the $yz$ electrons show the exponential decay more or less similar to the weak-excitation regime, with the initial maximum at ~250 fs. These contrastive behaviors solely depend on the orbital characters, not on experimental configuration. (Supplementary Note 4 and 5) To discuss the retardation behavior, the $F$ dependence of $\Delta I(t)$ for $xz$ is shown in Fig. 3i. In the weak-excitation regime ($F < F_c$), $\Delta I(t)$ curves commonly show the simple relaxation with the maximum around 250 fs. With increasing $F$, the retardation suddenly shows up at $F \approx F_c$. Its timescale estimated by $t_{\text{ret}}$ is 700 fs at $F \approx F_c$ and gradually decreases to 350 fs by increasing $F$ to 430 μJcm$^{-2}$.

**Fluence-dependent dynamics of electronic nematicity.**

Here we summarize the dynamics of the electronic nematicity in Fig. 4b. By fitting $\Delta k_{Fy}(t)$ curves in Fig. 2d (Supplementary Note 6), we plot $\tau_l$ and $t_p/2$ for respective $F$. $\tau_l$ is the exponential decay time indicating the quick recovery of nematicity. It shows a constant value (~800 fs) in the weak-excitation regime and a rapid decrease above $F > F_c$. The timescale of the Fermi surface oscillation indicated by $t_p/2$, which only appears in $F > F_c$, also rapidly decreases from 700 fs to 300 fs as increasing $F$. It shows that the oscillation gets more severely damped and hard to observe at high $F$. We also overlay the timescale of the retarded maximum in $xz$ component $t_{ret}$. As a result, $\tau_1$, $t_p/2$ and $t_{ret}$ similarly show the maximum values at $F_c = 220$ $\mu Jcm^{-2}$ that monotonically decrease with increasing $F$, while keeping the common relation $\tau_1 \approx t_{ret} \approx t_p/2$. The relation $t_{ret} \approx t_p/2$ implies that the orbital-dependent carrier dynamics is synchronized with the short-life nematic Fermi surface oscillation. We note that the transient Fermi surface at $t_p/2$ ($\approx t_{ret}$) is more elliptical than that expected without the oscillatory response. Such an overshoot of the nematicity in Fermi surface should also appear in the orbital-dependent carrier dynamics. In the process relaxing back from $C_4$ isotropic to $C_2$ nematic ground state, the electrons at the band top (black rectangle in Fig. 1 b,d) change their orbital characters from "(nearly) $xz/yz$ degenerate" to "predominantly $xz$". The retarded maximum in $I(t)$ for $xz$ can be thus regarded as an indication of the orbital redistribution from $yz$ to $xz$ (Fig. 4a). The synchronized responses in the Pomenranchuk Fermi surface oscillation and orbital-dependent carrier dynamics thus represent the nematic-orbital excitation.

**Discussion**

Now we discuss the nematic-orbital excitation obtained in the present TARPES by comparing with the nematic dynamics in thermal equilibrium as probed by the recent Raman scattering measurements.[15,31] The electronic Raman spectra of $XY$ symmetry ($X$ and $Y$ are coordinates along the crystal axes of the tetragonal setting) show the characteristic quasi-elastic peak (QEP) evolving toward $T_s$ on cooling the temperature ($T$), discussed in terms of nematic susceptibility enhancement.[15,31] The QEP rapidly diminishes at $T < T_s$, on the other hand, and a gap opens in the $XY$ Raman spectra thus indicating the suppression of low-energy nematic excitations (Ref. 31). These behaviors are reminiscent of the nematic-orbital excitation observed by TARPES, where the peculiar slowing down behavior shows up in $F > F_c$, and the excitation itself suddenly disappears in $F < F_c$. The nematic fluctuation is incoherent in nature, however, by instantaneously triggering the dissolution of the nematic state, it may be appearing as the heavily-damped oscillatory response in the time domain of non-equilibrium.

Further insight of the peculiar $F$ dependence can be obtained by plotting $t_p^{-1}$ and $\tau_1^{-1}$ (Fig. 4c). These values show a more or less $F$-linear behavior at $F > F_c$, indicating the critical slowing down. At $F = F_c$, $t_p^{-1}$ decreases down to 3.1 meV. In $F < F_c$, as already mentioned, the $k_F$ oscillation as well as the anomaly in the $xz$ orbital response disappear, and $\tau_1^{-1}$ becomes constant. In the $XY$ Raman spectrum, the $T$-linear behavior was found in the inverse of the QEP intensity above $T_s$ (Ref. 15), indicative of the Gaussian fluctuation evolving in this regime. Similarly, the elastoresistivity measurement had also revealed the existence of electronic nematic fluctuation at $T > T_s$ interpreted as the Curie-Weiss-like nematic susceptibility.[31] Through the analysis of the $T$-dependent nematic susceptibility in the form of $|T - T_0|^{-1}$, the authors discuss the mean-field transition temperature $T_0$ in terms of the ideal nematic transition purely driven by electrons without any influence of lattice.[15,31] For FeSe, $T_0$ was estimated to be far below $T_s$, *i.e.* 8 K, 20 K (Ref.15) and 30 K (Ref.32). The Curie-Weiss-like behavior of $t_p^{-1}$ and $\tau_1^{-1}$ toward $F \approx 40 \pm 20$ μJcm$^{-2}$, *i.e.* much smaller than $F_c = 220$ μJcm$^{-2}$, may be reflecting that the base temperature of the TARPES measurements (20 K) is close to $T_0$. This scenario is also consistent with the initial photo-response of $\Delta k_{Fy}$ with small threshold (< 30 μJcm$^{-2}$, see Fig. 2**c**). These results indicate that the electronic nematiciy in the initial ultrafast regime (~120 fs) shows the flexible photo-reaction by decoupling from the lattice. Our analysis on the transient electronic temperature ($T_e$) (Supplementary Note 7) indeed shows that $T_e$ immediately reaches $88 \pm 2$ K at 120 fs and then decreases in less than 1 ps (Supplementary Figure 6). For $t > 3$ ps, it becomes nearly constant at ~45 K, indicating the realization of quasi-equilibrium state where the temperatures of electrons and lattice become equivalent through the electron-lattice coupling[33].

The maximum lattice temperature is thus much lower than $T_s$ (= 90 K), showing that the lattice stays orthorhombic. We also note that the reduction of the lattice orthorhombicity is known to occur in a much slower time scale (e.g. ~30 ps) with a much higher pump fluence (e.g. 3.3 mJcm$^{-2}$) for BaFe$_2$As$_2$ (Ref.34).

The present results show that the femtosecond photon pulse can perturb the electronic nematic order and instantly decouple it from the lattice. Only in the strong-excitation regime where the nematic state is completely destroyed, there appears the peculiar dynamical process involving the orbital redistribution and short-life Pomenranchuk-type Fermi surface oscillation. This behavior is seemingly related to the critical nematic fluctuation, nevertheless, future theoretical studies on non-equilibrium critical phenomena are highly necessary. The recovery timescale of the nematic Fermi surface is strongly correlated with the short-life $k_F$ oscillation ($\tau_1 \approx t_p/2$), which also awaits investigations on the dynamics of fluctuation and dissipation in non-equilibrium states. Experimentally, further studies on the nematic dynamics around the quantum critical point in FeSe$_{1-x}$S$_x$ system[32] and the coherent nematic resonance mode in the superconducting state are highly desired. Systematic time-resolved diffraction measurements will also help understanding the possible interplay among the nematic excitation and the optical / acoustic phonons.[26, 35] The ultrafast photo-excitation adds a new possibility of understanding and manipulating the large-amplitude electronic fluctuations associated with unprecedented phenomena such as exotic superconductivity, peculiar magnetism, thermopower enhancement, and so on.

**Methods**

**Sample preparations.**

High-quality single crystals of FeSe were grown by the vapor transport method. A mixture of Fe and Se powders was sealed in an evacuated SiO$_2$ ampoule together with KCl and AlCl$_3$ powders[16]. The transition temperatures of the single crystals were estimated to be $T_s$ = 90 K and $T_c$ = 9 K from the electrical resistivity measurements. We showed the data obtained from five single crystals of FeSe which were synthesized less than two months before the TARPES measurements.

**Time and angle-resolved photoemission measurements.**

The TARPES measurements were done at ISSP, the University of Tokyo[28]. The laser pulse (1.5 eV and 170 fs duration) delivered from a Ti:Sapphire laser system operating at 250 kHz repetition (Coherent RegA 9000) was split into two branches: One is used as a pump, and the other was up-converted into 5.9 eV and used as a probe to generate photoelectrons. The delay origin $t = 0$ ps and time resolution (250 fs) were determined from the pump-probe photoemission signal of graphite attached near the sample. The photoelectrons were collected by a VG Scienta R4000 electron analyzer. The $E_F$ and the energy resolution (20 meV) were determined by recording the Fermi cutoff of gold in electrical contact to the sample. To detwin the single crystals, we applied an in-plane uniaxial tensile strain[17,19], which brings the orthorhombic $a$ axis ($a > b$) along its direction below $T_s$. We chose $s$ and $p$ polarizations to separately observe the $xz$ and $yz$ orbital electrons (see supplementary Note 3 for the details of experimental geometry and selection rule). Samples were cleaved *in situ* at room temperature in an ultrahigh vacuum of $5 \times 10^{-11}$ Torr.

**Data availability**

The datasets generated during and/or analyzed during the current study are available from the corresponding author on reasonable request.

## Acknowledgments

**General:** We thank M. Imada and Y. Yamaji for valuable discussions. We acknowledge H. Kontani and Y. Yamakawa for valuable discussions and band calculations. **Funding:** This research was supported by the Photon Frontier Network Program of the MEXT, the CREST project of the JST (Grant Number

JPMJCR16F2) and Grant-in-Aid for Scientific Research from JSPS, Japan (KAKENHI Grant No. 15H03687 and 18H01148).

**Author contributions**

T.Shim., and K.I. designed the research. T.S., Y.S., A.N., N.M. and Y.I. performed the TARPES measurements and analyzed the data. S.K., T.Shib., and Y.M. synthesized the single crystals. Y.I., and S.S. set up the TARPES apparatus. T.Shim. wrote the paper with inputs from T.Shib., Y.M., Y.I., S.S., and K.I..

**Competing interests**

The authors declare no competing interests.

**Figures and Tables:**

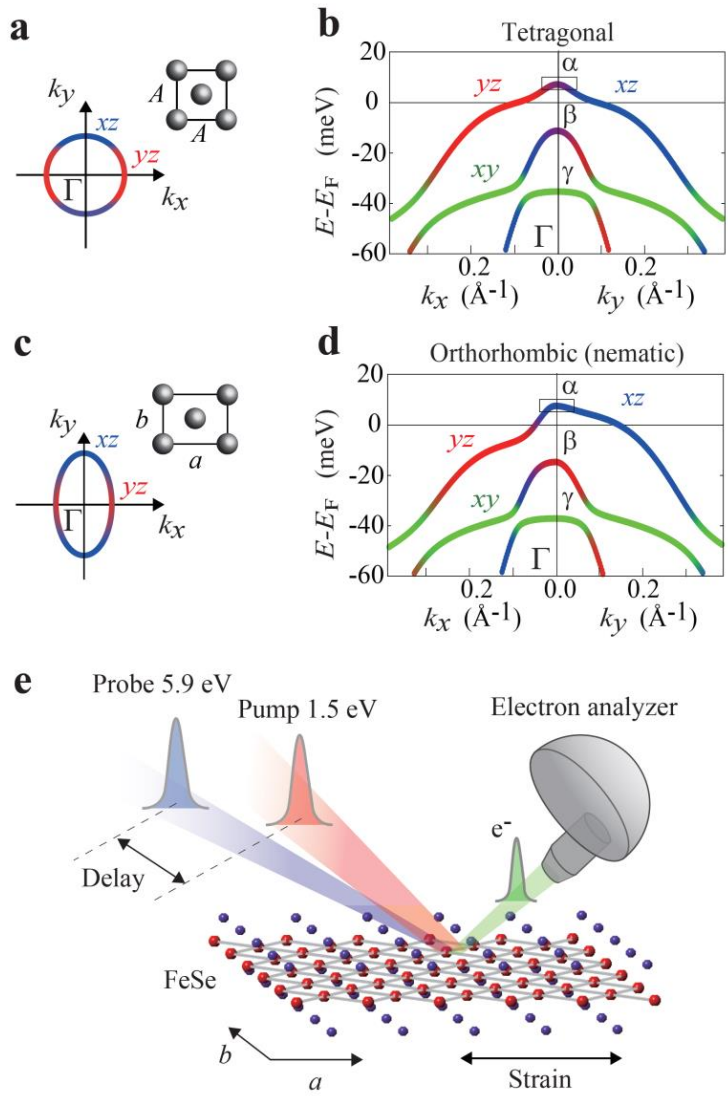

**Fig. 1. Electronic structure of FeSe and experimental setup for TARPES. a,** Schematic Fe lattice and Fermi surface around the Γ point in the tetragonal phase. $x$ and $y$ are coordinates along the crystal axes of the orthorhombic setting $a$ and $b$ ($a > b$), respectively. **b**, Band dispersions and orbital characters in the tetragonal phase obtained by a band calculation including the spin-orbit coupling[17]. α, β and γ denote the outer, middle and inner hole band, respectively. **c,d,** The same as **a** and **b** but for the orthorhombic phase. For reproducing the ARPES results (Supplementary Note 1), the spin-orbit coupling and orbital polarization were included in the band calculations in **d** (Ref.17). **e**, Schematic experimental geometry of TARPES on detwinned bulk FeSe.

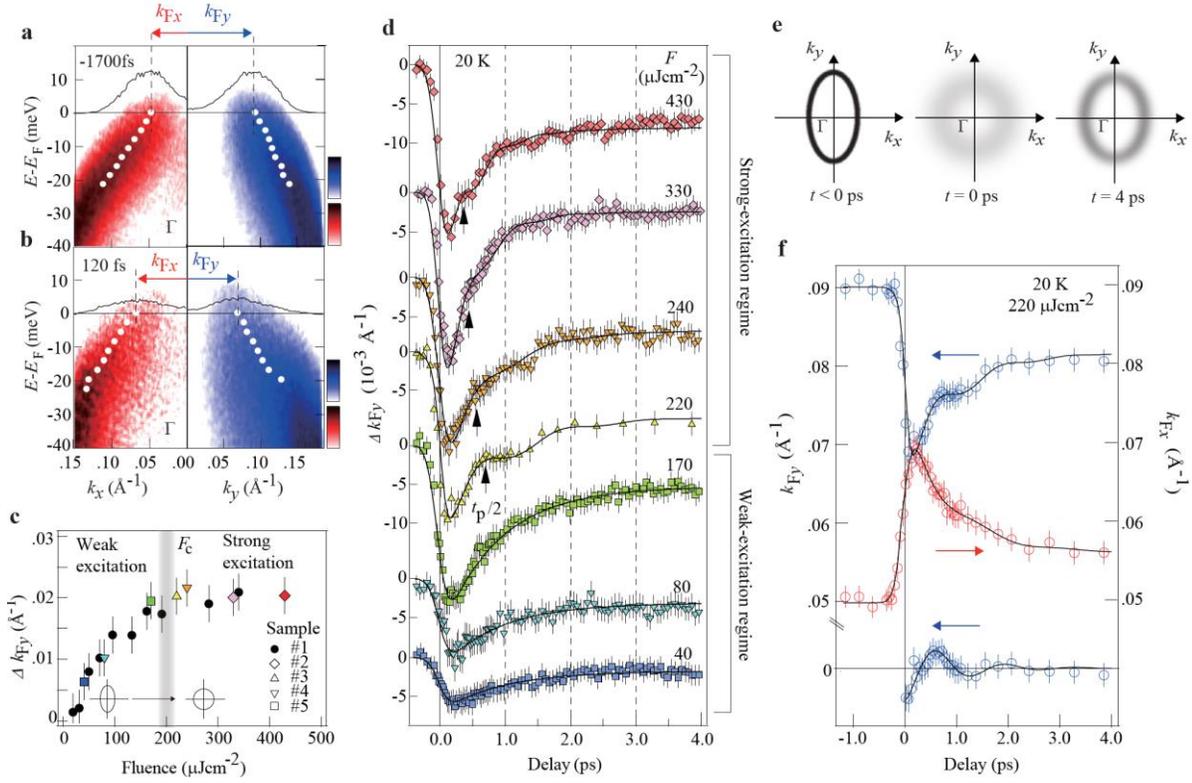

**Fig. 2. Ultrafast transformation of the nematic Fermi surface. a**, $E$-$k$ images in a logarithmic color scale obtained by $p$-polarized probe laser of $h\nu = 5.9$ eV along $k_x$ (left panel) and $k_y$ (right panel) axes at 20 K, before photo-excitation of $F = 220$ μJcm$^{-2}$ ($t = -1700$ fs). Black curves represent the momentum-distribution curves (MDCs) at the $E_F$ and the broken black lines indicate their peak positions. White markers show the band dispersions obtained from the MDC peaks. **b**, The same as **a** but after photo-excitation ($t = 120$ fs). **c**, $F$ dependence of the $k_F$ shift along $k_y$ ($\Delta k_{Fy}$) at $t = 120$ fs. $F_c$ represents the $F$ where the $C_4$ symmetric Fermi surface is attained after the photo-excitation. **d**, $F$ dependence of $\Delta k_{Fy}(t)$. The data set was obtained at 20 K from five single crystals as indicated by the different markers in **c**. **e**, Schematics of the Fermi surface around the Γ point for $t < 0$ ps, $t = 0$ ps and $t = 4$ ps deduced from the transient $k_{Fx}$ and $k_{Fy}$, and the width of the MDCs at the $E_F$. **f**, Transient $k_{Fy}$ (blue open circles) and $k_{Fx}$ (red open circles) as a function of delay time with the fitting functions (black curves). Damped oscillation in $k_{Fy}$ was extracted by subtracting the exponential decay components, as shown in the bottom.

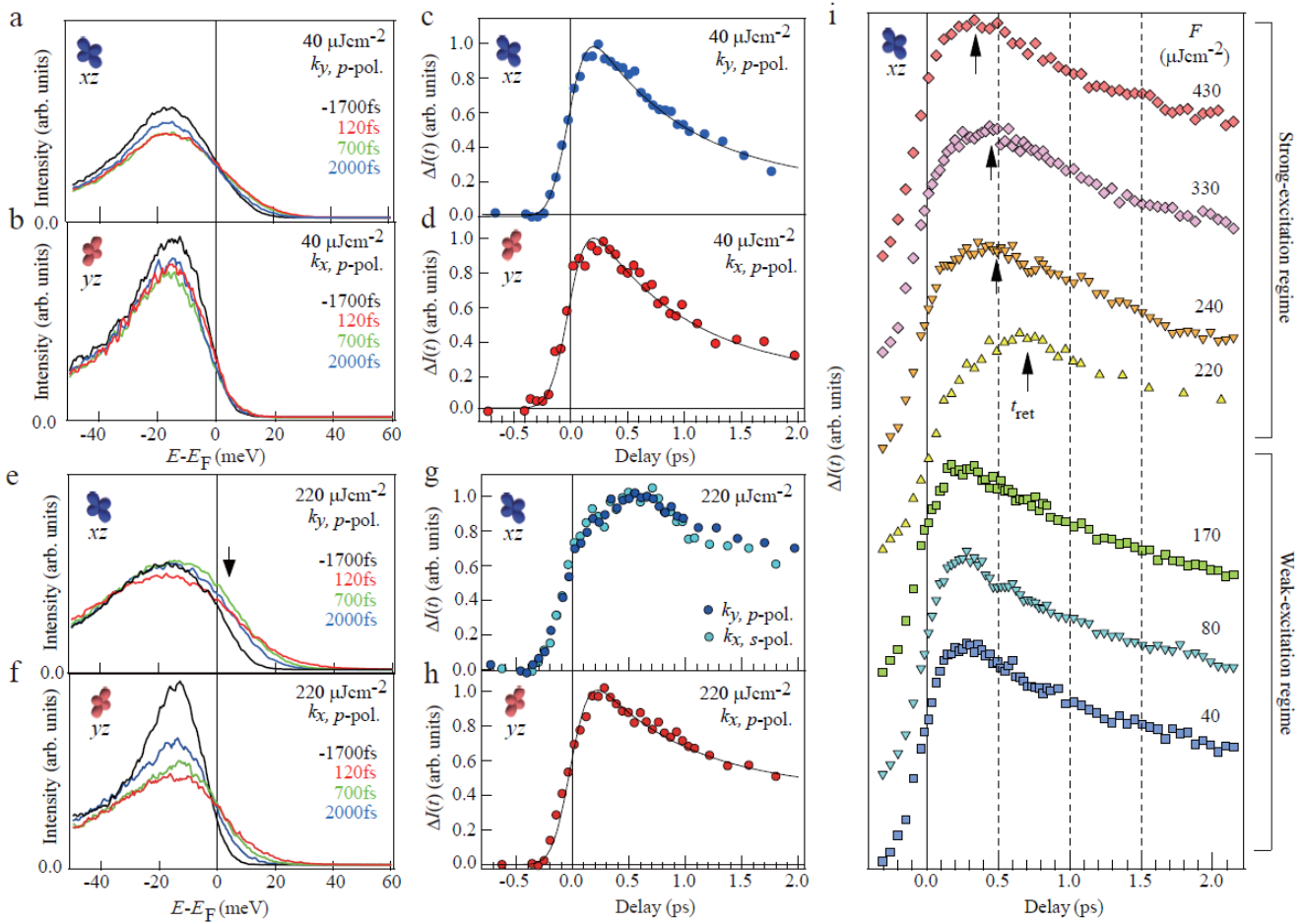

**Fig. 3. Orbital-dependent dynamics of the photo-excited electrons at the Γ point. a**, EDCs around the Γ point ($k = 0.0 \pm 0.04$ Å$^{-1}$) obtained from the $E$-$k$ images at 20 K along $k_y$ at $t = -1700$ fs, 120 fs, 700 fs and 2000 fs, in the $p$-polarized setting. The fluence of the photo-excitation was $F = 40$ μJcm$^{-2}$. **b**, The same as **a** but obtained along $k_x$ axis. $xz$ ($yz$) electrons are dominantly observed in **a** (**b**). **c,** The photo-electron intensities normalized by that before photo-excitation, $\Delta I(t)$, at the α band top ($E - E_F = 7.5 \pm 2.5$ meV) around the Γ point ($k_y = 0.0 \pm 0.04$ Å$^{-1}$), obtained by the $p$-polarized probe laser and the pump fluence of $F = 40$ μJcm$^{-2}$. The black curve represents the fitting function assuming the double exponential components with the time constants of 850 fs and 80 ps. **d**, The same as **c** but at $k_x = 0.0 \pm 0.04$ Å$^{-1}$. **e-h**, The same as **a** - **d** but with the pump fluence of $F = 220$ μJcm$^{-2}$. The light blue markers in **g** represent the $\Delta I(t)$ of $xz$ obtained by $s$-polarized probe laser. (Supplementary Note 4) **i**, $F$ dependence of the normalized $\Delta I(t)$ for $xz$. The black arrows represent the time of the retarded maxima $t_{\text{ret}}$ for $F > F_c$.

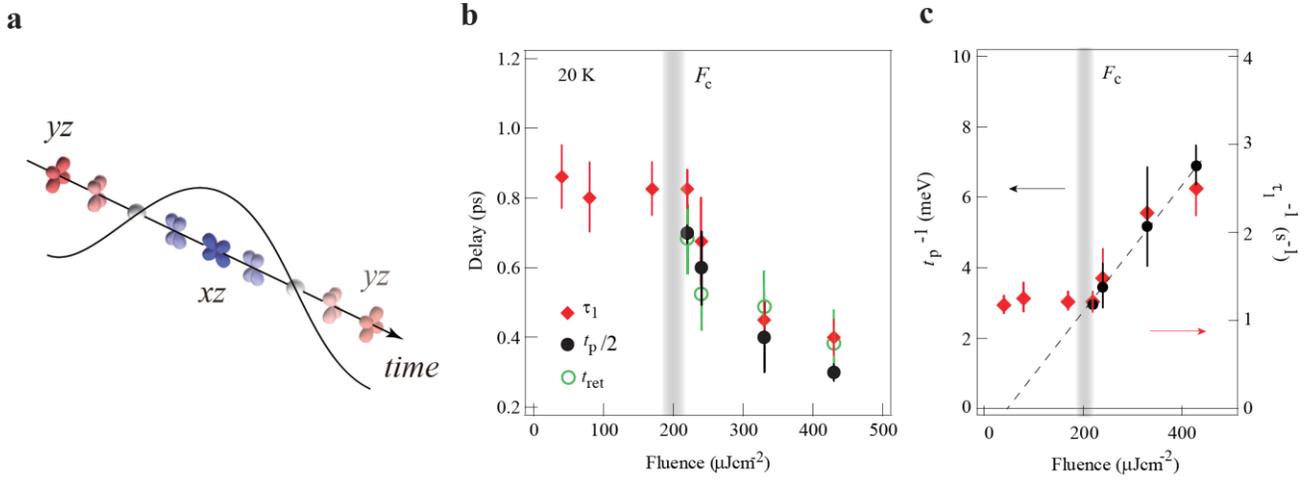

**Fig. 4. Pump-fluence dependence of the time scales for the orbital and nematicity dynamics. a,** Illustration of the orbital excitation in FeSe. **b,** $F$ dependences of $t_{ret}$ (open green circles), $t_p/2$ (filled black circles) and $\tau_1$ (filled red diamonds). **c,** $t_p^{-1}$ (filled black circles) and $\tau_1^{-1}$ (filled red diamonds) plotted as functions of $F$. The dotted line is a linear function derived from the data at $F > F_c$.

# Supplementary Information for

# Ultrafast nematic-orbital excitation in FeSe

T. Shimojima[1,2], Y. Suzuki[2], A. Nakamura[2], N. Mitsuishi[2], S. Kasahara[3], T. Shibauchi[4], Y. Matsuda[3], Y. Ishida[5], S. Shin[5] and K. Ishizaka[1,2]

[1]RIKEN Center for Emergent Matter Science (CEMS), Wako 351-0198, Japan

[2]Quantum-Phase Electronics Center (QPEC) and Department of Applied Physics, The University of Tokyo, Tokyo 113-8656, Japan

[3]Department of Physics, Kyoto University, Kyoto 606-8502, Japan

[4]Department of Advanced Materials Science, The University of Tokyo, Kashiwa, 277-8561, Japan

[5]Institute for Solid State Physics (ISSP), The University of Tokyo, Kashiwa, 277-8581, Japan.

**Supplementary Note 1: Laser ARPES on detwinned FeSe in thermal equilibrium**

Here we show the ARPES data[1] which was used to draw the schematic band structures in Fig. 1**b** and **d**. Supplementary Figure 1**a** and **b** show the experimental geometries used for the laser-ARPES measurements in thermal equilibrium. In the geometry in Supplementary Fig. 1**a** (1**b**), the strain direction is parallel (perpendicular) to the detector slit and the momentum along $k_x$ ($k_y$) is measured. Here, $xz$ ($yz$ and $xy$) orbitals have even (odd) symmetry with respect to the mirror plane and can be detected by $p$ ($s$)-polarized probe laser. Similarly, $yz$ ($xz$ and $xy$) are detected by $p$ ($s$)-polarized laser in the geometry in Supplementary Fig. 1**b** which measures along $k_y$. The observable orbital characters are also indicated. Here, we can focus mainly on the $xz$ and $yz$ orbitals, since the photoelectron intensity of the $xy$ orbital is quite weak near $E_F$ in the ARPES measurements on FeSe[1,2].

In the tetragonal phase, FeSe exhibits a nearly circular Fermi surface around the Γ point as depicted in Fig. 1**a**. Along $k_x$ ($k_y$), the band forming the Fermi surface is known to have the $yz$ ($xz$) orbital component, which can be confirmed by the contrast of the polarization-dependent ARPES intensity indicated in the square boxes in Supplementary Fig. 1**c** and **d**. In the orthorhombic (nematic) phase, on the other hand, the momentum-dependent orbital polarization[1] modifies the shape of the Fermi surface into an elliptical one as shown in Fig. 1**c**, and the $k_{Fx}$ and $k_{Fy}$ become inequivalent as indicated by the black double-headed arrows in Supplementary Fig. 1**e** and **f**. At the same time, the light polarization dependence inside the square boxes now shows that the orbital components at the $k_F$'s are mixed, especially along $k_x$, the shorter axis (Supplementary Fig. 1**e**). The band dispersions and the orbital characters were reproduced by the calculations including both the orbital non-equivalency and the spin-orbit coupling[1] as shown in Fig. 1**d**.

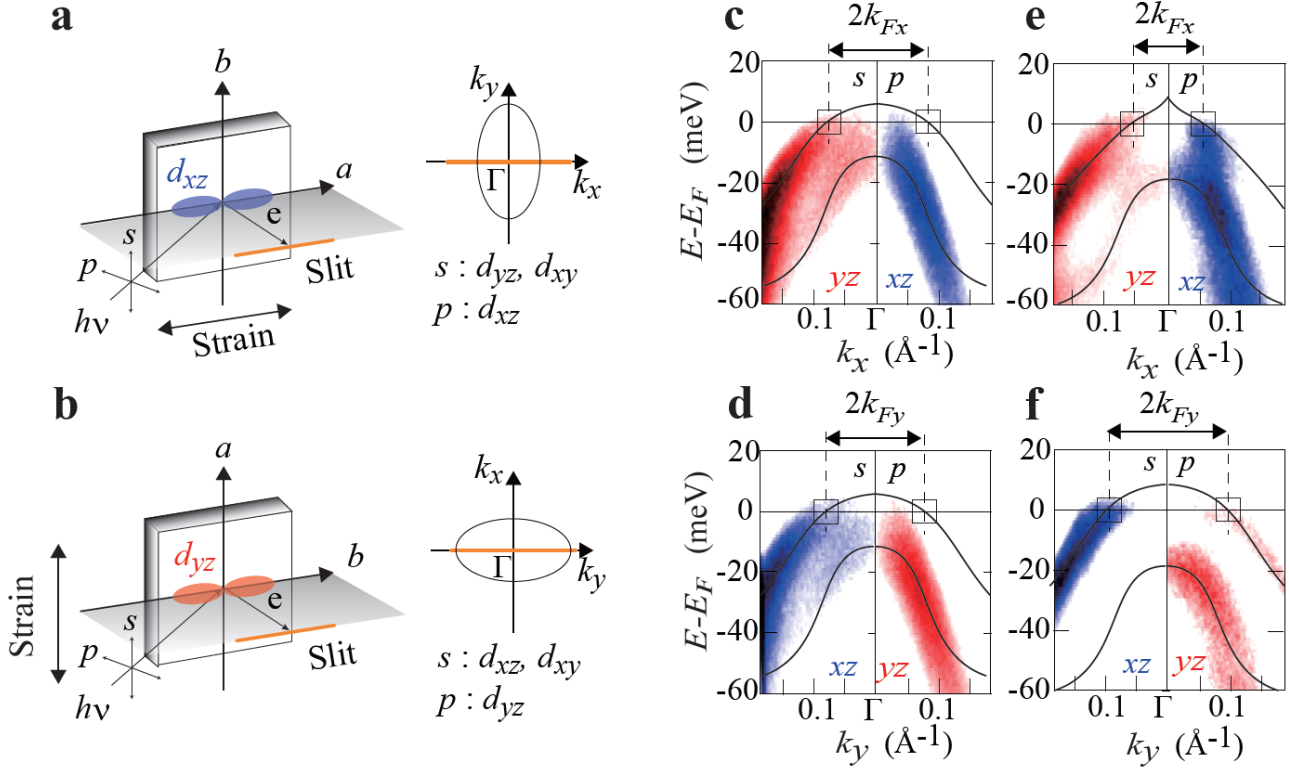

**Supplementary Fig. 1. Experimental geometries for ARPES and *E-k* images of detwinned FeSe. a,b,** The experimental geometries for polarization-dependent laser ARPES[1] in thermal equilibrium. In the geometry in **a** (**b**), orthorhombic *a* (*b*) axis is parallel to the detector slit. Gray plane represents a mirror plane of the orthorhombic lattice. We used the linear polarization *s* (*p*) which is perpendicular (parallel) to the detector slit. Momentum cut is shown in each geometry with the orange line. Observable orbital characters are also indicated for each polarization. **c**, Band dispersions along $k_x$ axis around the Γ point of the detwinned FeSe obtained at 160 K with the *s*- and *p*-polarized laser ($h\nu$ = 5.9 eV). Black curves represent hole band dispersions. Dotted lines highlight the position of $k_F$. **d**, The same as **c** but along $k_y$. **e,f,** The same as **c** and **d** but obtained at 30 K.

**Supplementary Note 2: Momentum dependence of the orbital characters in the hole Fermi surface around the Γ point**

Here we discuss the momentum dependence of the orbital characters in the hole Fermi surface for understanding the redistribution of the orbital components across the nematic order. Supplementary Figures 2**a** and **b** are the Fermi surface-angle ($\theta$) dependence of the orbital components in the tetragonal and orthorhombic phases, respectively, obtained from the calculations based on the *d-p* model[1] assuming the spin-orbit coupling and orbital non-equivalency, which exhibits the Fermi surface shape similar to that obtained by previous laser ARPES[1]. While the orbital components other than *xz* and *yz* orbitals were also mixed (~20 %) in the hole Fermi surface both below and above the nematic order, we focus on *xz* and *yz* orbitals for discussing the temperature dependence related to the nematic order.

In the tetragonal phase (Supplementary Fig. 2**a**), the band forming the Fermi surface has the dominant contribution of *yz* (*xz*) orbital along $k_x$ ($k_y$). Here, *k*-dependent orbital characters keep the $C_4$ symmetry. In the orthorhombic phase (Supplementary Fig. 2**b**), the elliptical Fermi surface is composed of both *xz* and *yz* orbitals along $k_x$ axis due to the orbital polarization, while that along $k_y$ axis exhibits almost *xz* character. These $\theta$ dependences of the orbital components are qualitatively consistent with the polarization-dependent ARPES results shown in Supplementary Fig. 1**c** - **f**.

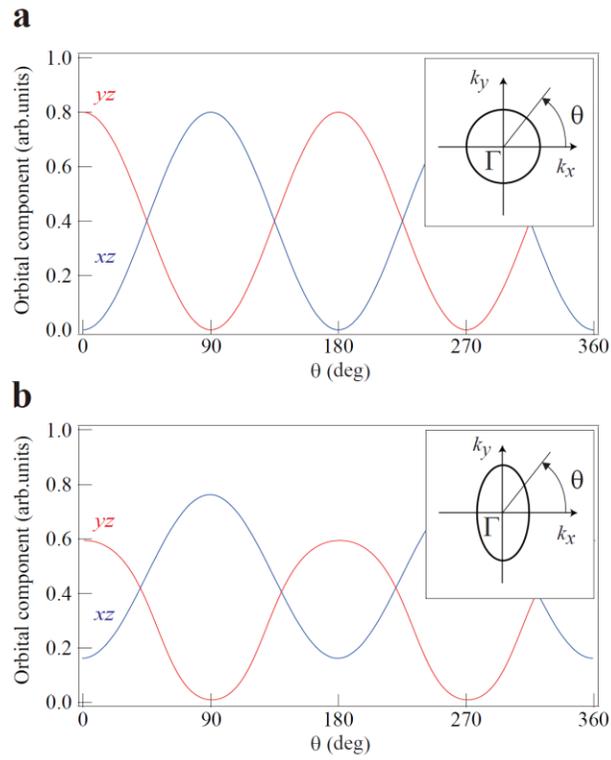

**Supplementary Fig. 2. Calculated orbital characters of the Fermi surface around the Γ point of FeSe.**
**a,b,** The Fermi surface-angle ($\theta$) dependence of the orbital components in the tetragonal and orthorhombic phases, respectively. The blue (red) curves represent orbital component for the *xz* (*yz*) orbital. The inset shows the definition of the Fermi surface angle $\theta$.

**Supplementary Note 3: Experimental geometries for TARPES on detwinned FeSe**

In Supplementary Figure 3**a**, we show the detwinning device made for TARPES measurements. Single crystals (2 mm × 4 mm × 0.05 mm) were cut along the orthorhombic axes and pulled along one of the axes by tightening the screw. The longer *a* axis is then aligned parallel to the direction of the tensile strain below $T_s$. Supplementary Figures 3**b** and **c** show the experimental geometries used for the TARPES measurements. In the geometry in Supplementary Fig. 3**b** (3**c**), the strain direction is set parallel (perpendicular) to the detector slit and the momentum along $k_x$ ($k_y$) is measured. Here, we consider a mirror plane spanned by the surface normal and the detector slit. When using the *s*-polarized light, only the orbitals of even parity with respect to the mirror plane, i.e. *xz*, is observed due to the selection rule. Similarly, only *yz* is detected by the *s*-polarization in the geometry in Supplementary Fig. 3**c** which measures along $k_y$. Note that the selection rule for *p*-polarization holds only at the Γ point (*yz* for Supplementary Fig. 3**b** and *xz* for Supplementary Fig. 3**c**), by considering the other mirror plane spanned by the surface normal and the orthorhombic axis perpendicular to the detector slit. However, we can confirm that the orbital-selective observation remains capable also away from the Γ point, by directly comparing with the *E-k* images obtained by another laser-ARPES ($h\nu$ = 5.9 eV) system with a higher symmetry configuration (Supplementary Note 1). Owing to the orbital selectivity, we can separately discuss the *xz* and *yz* orbitals in the present TARPES measurements on FeSe.

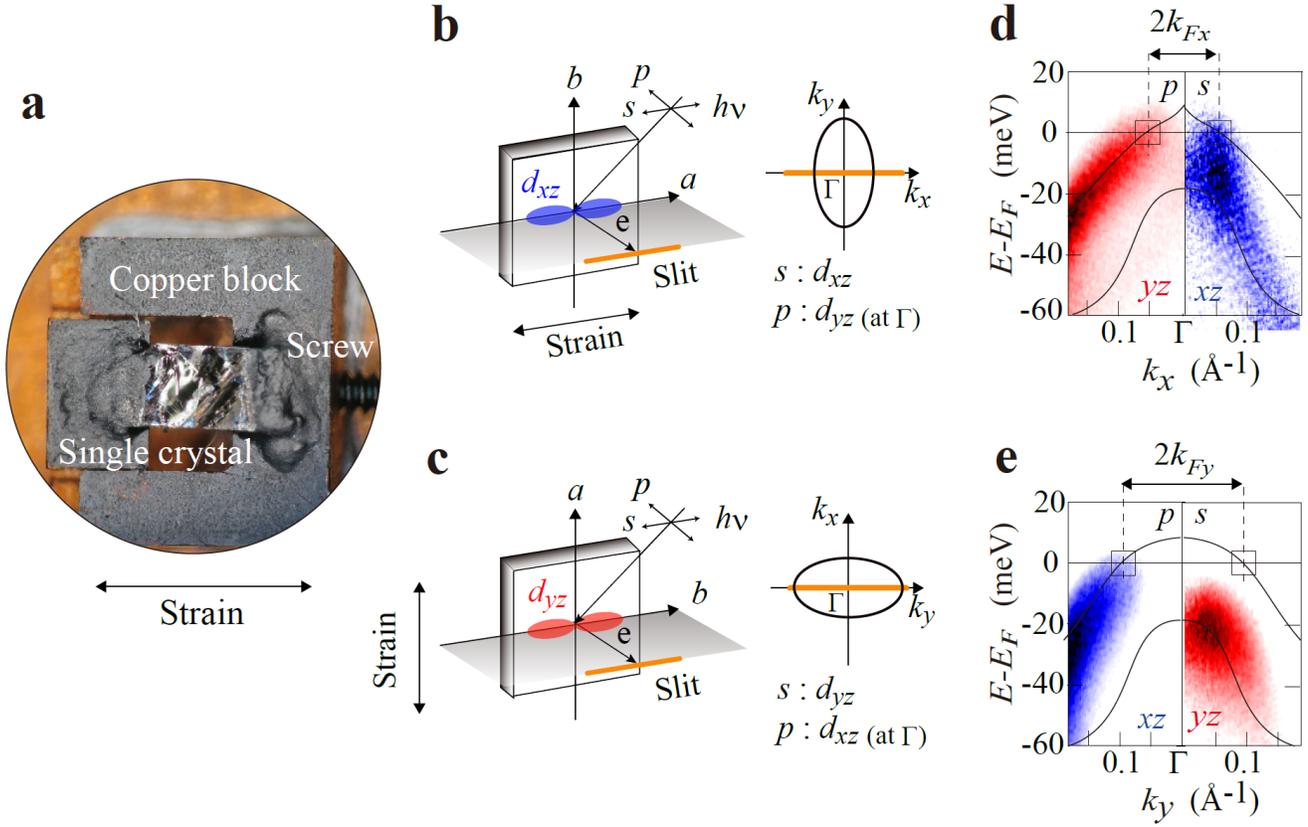

**Supplementary Fig. 3. Experimental geometries for TARPES and *E-k* images of detwinned FeSe before photo-excitation. a**, Detwinning device with the strained single crystal of FeSe. **b,c**, The experimental geometries for TARPES where the detector slit is parallel to the orthorhombic *a* and *b* axis, respectively. Gray plane represents a mirror plane of the orthorhombic lattice. The linear polarization *s* is parallel to the detector slit. Momentum cut is shown for each geometry with the orange line. Observable orbital characters are also indicated for each polarization. **d**, Band dispersions along $k_x$ axis around the Γ point of the detwinned FeSe obtained at 20 K with the *s*- and *p*-polarized probe laser ($h\nu = 5.9$ eV) in the geometry of **b**. Black curves represent hole band dispersions. Dotted lines highlight the position of $k_F$. **e**, The same as **d** but along $k_y$ in the geometry of **c**.

**Supplementary Note 4: Carrier relaxation dynamics for *xz* electrons obtained by different linear polarizations of probe laser.**

In Supplementary Figure 4**a, b,** we show the EDCs around the Γ point for *xz* electrons obtained by *p*-polarized probe laser along $k_y$, and *s*-polarized probe laser along $k_x$ in the strong-excitation regime ($F = 220$ μJcm$^{-2}$) at 20 K, respectively. In both cases, we observe a non-monotonic relaxation of photo-excited *xz* electrons ($E_F < E < 10$ meV) which keep increasing from $t = 120$ fs to 700 fs. The $\Delta I(t)$ for *xz* obtained by *s*-polarization is almost identical to that of *p*-polarization as shown in Fig. 3g. We thus confirm that the retarded maximum in $\Delta I(t)$ solely depends on the orbital character, not on the experimental configurations such as probe-laser polarizations (*p* or *s*) and probed momentum directions ($k_x$ or $k_y$).

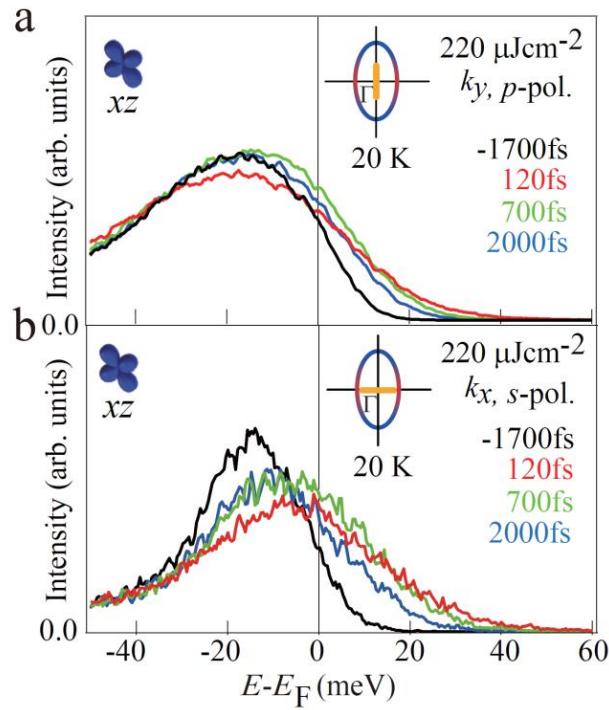

**Supplementary Fig. 4. The EDCs around the Γ point probed by different probe laser polarizations in the strong-excitation regime. a,** Time dependence of the EDCs around the Γ point ($k_y = 0.0 \pm 0.04$ Å$^{-1}$) for *xz* electrons obtained by *p*-polarized probe laser in the strong-excitation regime ($F = 220$ μJcm$^{-2}$) at 20 K. The orange bar in the inset shows the integrated momentum region for obtaining the EDCs. **b,** The same as **a** but obtained by *s*-polarized probe laser along $k_x$.

**Supplementary Note 5: Temperature dependence of the carrier relaxation dynamics for *xz* and *yz* electrons in the strong excitation regime.**

In Supplementary Figure 5**a, b,** we show the EDCs around the $\Gamma$ point ($k_y = 0.0 \pm 0.04$ Å$^{-1}$) for *xz* electrons obtained by *p*-polarized probe laser in the strong-excitation regime ($F = 220$ μJcm$^{-2}$) at 20 K and 160 K, respectively. The photo-excited tail intensity of EDCs ($E_F < E < 10$ meV) at 160 K monotonically increases after photo-excitation. The indication of the retarded maximum in $\Delta I(t)$ is absent in the tetragonal phase.

In Supplementary Figure 5**c, d**, we also show the EDCs around the $\Gamma$ point ($k_x = 0.0 \pm 0.04$ Å$^{-1}$) for *yz* electrons obtained by *p*-polarized probe laser in the strong-excitation regime ($F = 220$ μJcm$^{-2}$) at 20 K and 160 K, respectively. At 160 K, the line shape and time dependence of the EDCs for *yz* (Supplementary Fig. 5**d**) are almost comparable to those for *xz* (Supplementary Fig. 5**b**) reflecting the tetragonal symmetry. We thus confirm that the non-monotonic carrier relaxation for *xz* electrons appears only in the nematic phase.

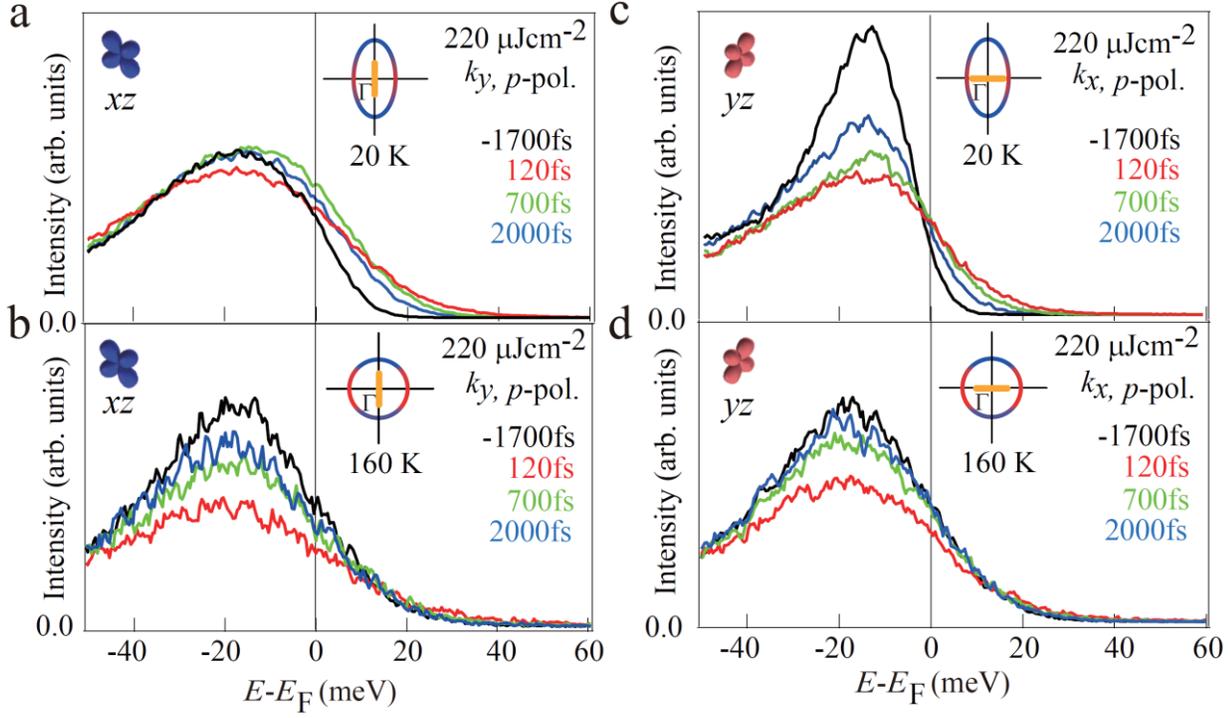

**Supplementary Fig. 5. Temperature dependence of the EDCs around the Γ point in the strong-excitation regime. a,** Time dependence of the EDCs around the Γ point ($k_y = 0.0 \pm 0.04$ Å$^{-1}$) for $xz$ electrons obtained by $p$-polarized probe laser in the strong-excitation regime ($F = 220$ μJcm$^{-2}$) at 20 K. **b,** The same as **a** but at 160 K. **c,d,** The same as **a** and **b** but for $yz$ electrons obtained along $k_x$.

**Supplementary Note 6: Fluence dependence of the time constants obtained from the fitting procedures on the $k_{Fy}(t)$ curves.**

We show the time constants obtained from the fitting analysis on the $k_{Fy}(t)$ for several $F$ values assuming the function $k_F(t) = k_{F1}\exp(-t/\tau_1) + k_{F2}\exp(-t/\tau_2) + k_{F3}\exp(-t/\tau_3)\cos(2\pi t/t_p)$ as summarized in the Supplementary Table 1. For $F = 40$, 80 and 170 µJcm$^{-2}$, we assumed the double exponential components.

| F (µJcm$^{-2}$) | $\tau_1$ (ps) | $\tau_2$ (ps) | $\tau_3$ (ps) | $t_p$ (ps) |
|---|---|---|---|---|
| 430 | 0.40 | ≥ 80 | 0.40 | 0.6 |
| 330 | 0.45 | ≥ 80 | 0.55 | 0.8 |
| 240 | 0.68 | ≥ 80 | 0.60 | 1.2 |
| 220 | 0.83 | ≥ 80 | 0.55 | 1.4 |
| 170 | 0.83 | ≥ 80 | - | - |
| 80 | 0.8 | ≥ 80 | - | - |
| 40 | 0.85 | ≥ 80 | - | - |

**Supplementary Table 1.** Time constants extracted from the $k_{Fy}(t)$ curves.

**Supplementary Note 7: Estimation of the transient electronic temperature.**

Here we estimate the electronic temperature ($T_e$) from the fitting analysis of the momentum-integrated EDCs. In general, $T_e$ should be estimated by using the momentum-integrated EDC spectrum which represents the total density of states multiplied by the Fermi-Dirac function further convoluted by the instrumental resolution function. We integrated the EDCs of ARPES on $xz$ from 0.0 Å$^{-1}$ to 0.17 Å$^{-1}$ along $k_y$, and fitted by a FD function convoluted by the gaussian of energy resolution (20 meV), assuming the constant density of states neat $E_F$ (Supplementary Fig. 6**a**). After the photoexcitation of 220 μJcm$^{-2}$, $T_e$ reaches 88 ±2 K at 120 fs. Then, it shows a rapid decrease in < 1 ps and remains nearly constant at 45 K for $t > 3000$ fs (Supplementary Fig. 6**b**), which is considerably lower than $T_s = 90$ K.

According to the two-temperature model[3], elevated $T_e$ approaches a constant value after the rapid relaxation *via* the electron-lattice coupling. There, the quasi-equilibrium state is realized, where the temperatures of electrons and lattice become equivalent. This behavior has been indeed discussed in the ultrafast optical measurements of the iron-based superconductors[4]. The maximum lattice temperature is thus expected to be ~45 K in the present TARPES case. These analyses suggest that the electronic nematic order gets dissolved in the ultrafast regime while the lattice well maintains the orthorhombicity.

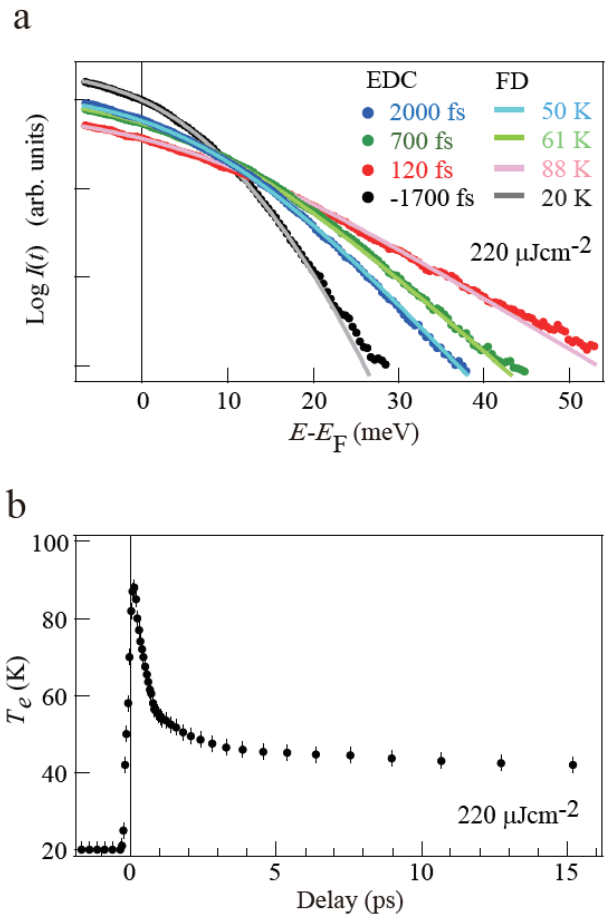

**Supplementary Fig. 6. Estimation of the electronic temperature. a**, Transient EDCs for *xz* and FD functions assuming a constant density of states. **b**, Time dependence of the electronic temperature for *xz* at 220 μJcm$^{-2}$.

**Supplementary References:**

[1] Suzuki, Y. *et al. Phys. Rev. B* **92**, 205117 (2015).

[2] Shimojima, T. *et al. Phys. Rev. B* **90**, 121111(R) (2014).

[3] Anisimov, S.I. *et al. J. Exp. Theor. Phys.* **66**, 375 (1974).

[4] Patz, A. *et al. Nature Communications* **5**, 3229 (2014).